\documentstyle[editedvolume,epsf]{crckapb} 

\begin{opening}
\title{The Mass Function at the End of the Main Sequence:\protect\\
 the M35 Open Cluster}

\author{D. Barrado y Navascu\'es}
\institute{Max-Planck Institut f\"ur Astronomie\\
K\"onisgtuhl 17, D-69117 Heidelberg, Germany}
\author{J.R. Stauffer}
\institute{Harvard-Smithsonian Center for Astrophysics\\
           60 Garden St. Cambridge, MA02138}
\author{J. Bouvier}
\institute{Observatoire de Grenoble\\
     F-38041 Grenoble, Cedex 9, France}
\author{E.L. Mart\'{\i}n}
\institute{University of California at Berkeley\\
          601 Cambell Hall, Berkeley, CA 94720, USA}

\end{opening}

\runningtitle{The Mass Function of M35}

\begin{document}

\section{Introduction}

The Open Cluster M35 (NGC~2168) is  
 moderately nearby, with (m-M)$_0$=9.7 (Vidal 1973). It is very well
 populated, with a total mass estimated between 1000 and 3000 M$_\odot$. 
Although the cluster is at relative low galactic latitude 
({\it l}$^{\sc \,II}$=186.58, {\it b}$^{\sc \,II}$=2.19),
 M35 is rich enough so that it is easy to 
separate the field stars and the cluster population.
 On the other hand, its reddening is not too high
 --E(B--V)=0.17, and its age  is very close to the 
Pleiades value ($\sim$125$\times$10$^6$ yr, Stauffer et al. 1998),
 since their turn-off point are in the same location (Vidal 1973).
We have carried out photometric observations of the cluster  
using the Canada--France-Hawaii Telescope 
 (December, 1996) and the new mosaic camera
 (8000$\times$8000 pixels),
covering a region of 30'$\times$30', equivalent to about
 a fourth of the total area of the cluster. 
 We collected data
in the magnitude interval 17.5 $\le$ I$_{\rm c}$ $\le$ 23.5,
  0.5 $\le$ (R-I)$_{\rm c}$ $\le$ 2.5. Errors can be estimated 
as 0.05 for each filter.  The survey is complete down to 
I$_{\rm c}$=21, (R-I)$_{\rm c}$=2.0. Details can be found in
Barrado y Navascu\'es et al. (1999).

\begin{figure}	
\vspace{-2.0cm}
\hbox{\hspace{-0.6cm}\epsfxsize=5.8cm \epsfbox{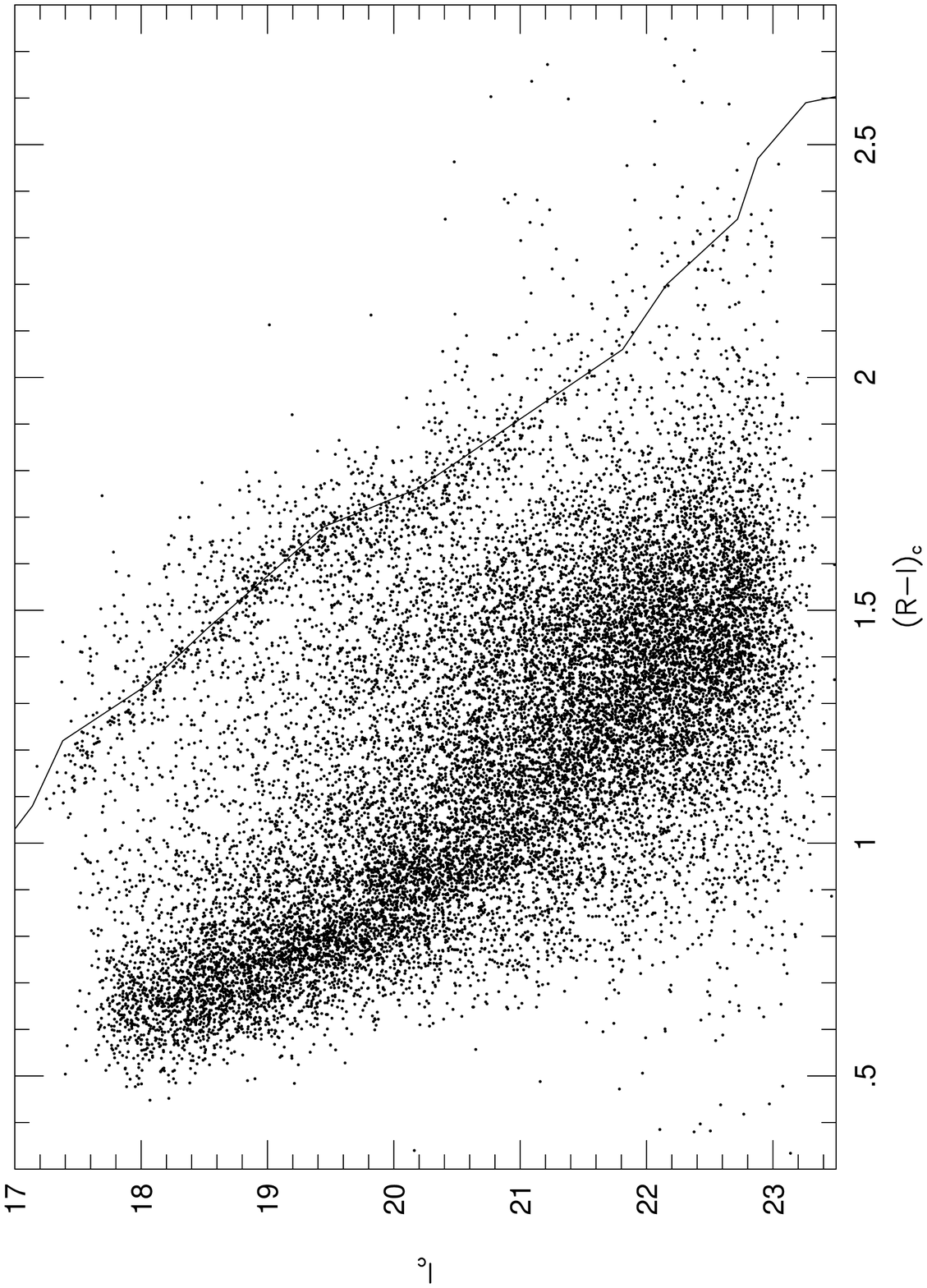}
\hspace{6.5mm}\epsfxsize=5.8cm \epsfbox{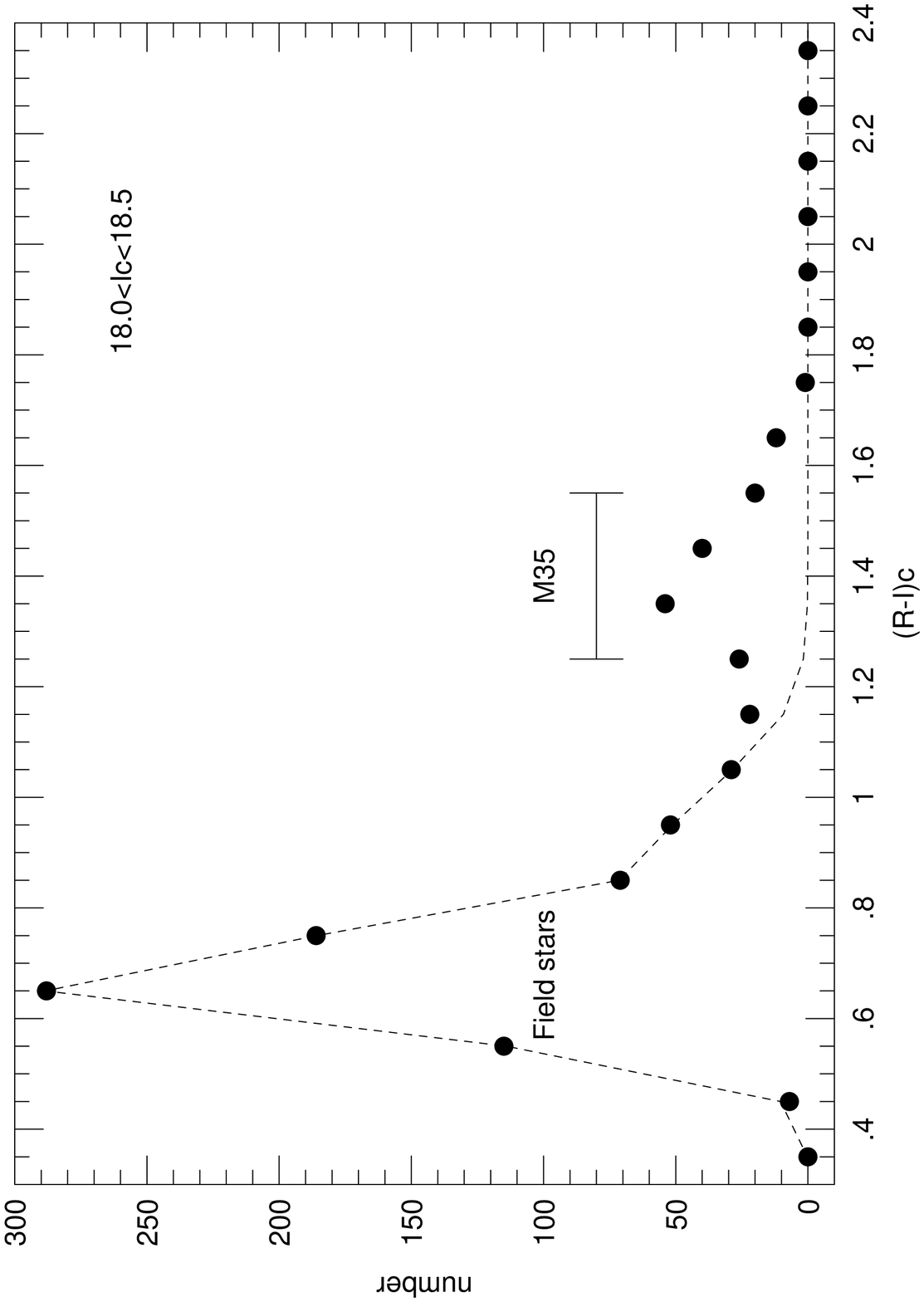}}
\vspace{0cm}
  \caption{
{\bf a} Color-Magnitude Diagram for M35.  
{\bf b} Histogram for one slice selected from the CMD.
The number of stars in each (R-I) bin is shown as filled circles, whereas
the fit for field stars appears as a dashed line.}
\end{figure}

\section{The selection of candidates}

We selected candidates to be members of the clusters 
based on the position of a Color-Magnitude diagram  (CMD).
Figure 1a shows  I$_{\rm c}$ versus (R--I)$_{\rm c}$. 
 We have added an empirical ZAMS
for field stars, derived from  Leggett (1992),
shifted with the reddening -E(R-I)${\rm c}$=0.17,- and the
distance module  --(m-M)$_{\rm I}$=10.05.--
of M35. Note that the ZAMS  accurately
traces the relative concentration of stars in M35, as it does with
Pleiades stars (Bouvier et al. 1998).
Figure 1a shows clearly two accumulations of stars: Most of the
detected stars lie between 0.5 $<$ (R--I)$_{\rm c}$ $<$ 1.7,
following a distribution parallel to the ZAMS, but 4 magnitudes
fainter. A significant fraction of the stars are on the locus of
M35. However, there is not a clear gap between the former and the
latter  groups. In any case, the diagram allows us to perform a
tentative identification of candidates for membership to the
cluster.  They were
selected based on the fact that the number of stars for a given
bin in magnitude decreases as the color increases for field
stars, except for the area close to the cluster, where a sudden
increase appears. Therefore, we selected those stars in  a band
around the relative maximum, taking into account the errors and 
the effect of binarity.

The relatively smooth transition between the concentration of
field stars, fainter and/or redder than M35, and this cluster
indicates that our list of candidates is strongly poluted
 by field stars. In order to estimate this contamination, 
we  have divided our CMD into strips
parallel to the (R--I)$_{\rm c}$ axis, 0.5$^{\rm mag}$ wide in 
I$_{\rm c}$. Each strip was subdivided into intervals of 
(R--I)$_{\rm c}$, 0.1$^{\rm mag}$ wide. Then, we counted the
number of stars in each interval or box, N$_{\rm *}$.
We fitted 3 Gaussian curves to the pairs 
[N$_{\rm *}$,(R--I)$_{\rm c}$] for a given I$_{\rm c}$ interval:
 one for the main population of field stars, another for the transition to
M35, and the last one for the members of M35. The first two allowed
us to estimate the contamination of field stars in the region
where M35 lies (see Figure 1b).
 For stars fainter than I$_{\rm c}$=22, we were
not able to obtain reliable fits due to the lack of
differentiation between field stars and the cluster, and,
therefore,  we cannot estimate the contamination by spurious
members. Note, however, that the brown dwarf domain would start
at I$_{\rm c}\sim$22.2, (R--I)$_{\rm c}\sim$2.3 (if M35 really is
the same age as the Pleiades).

\begin{figure}	
\vspace{-2.0cm}
\hbox{\hspace{-0.6cm}\epsfxsize=5.8cm \epsfbox{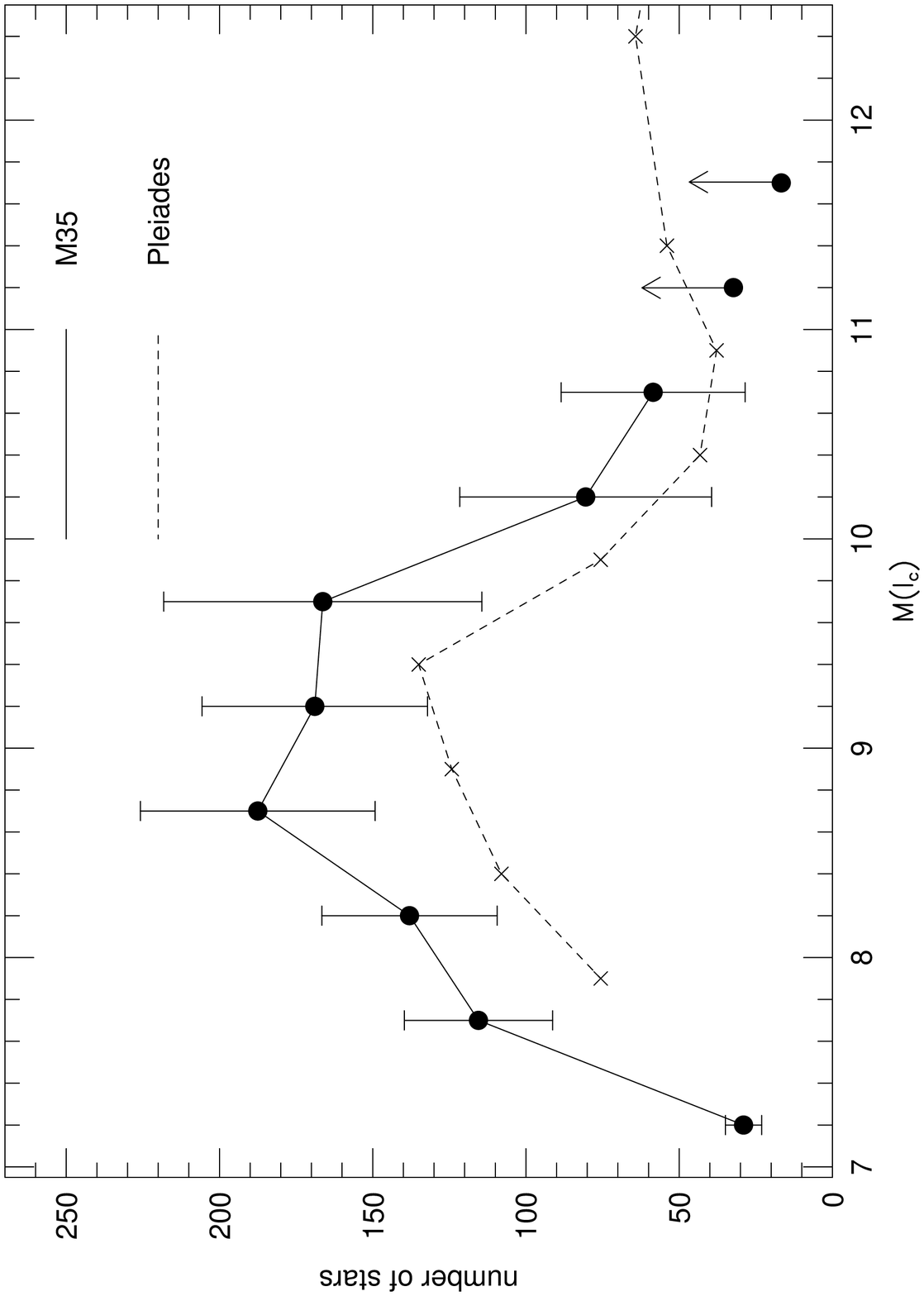}
\hspace{6.5mm}\epsfxsize=5.8cm \epsfbox{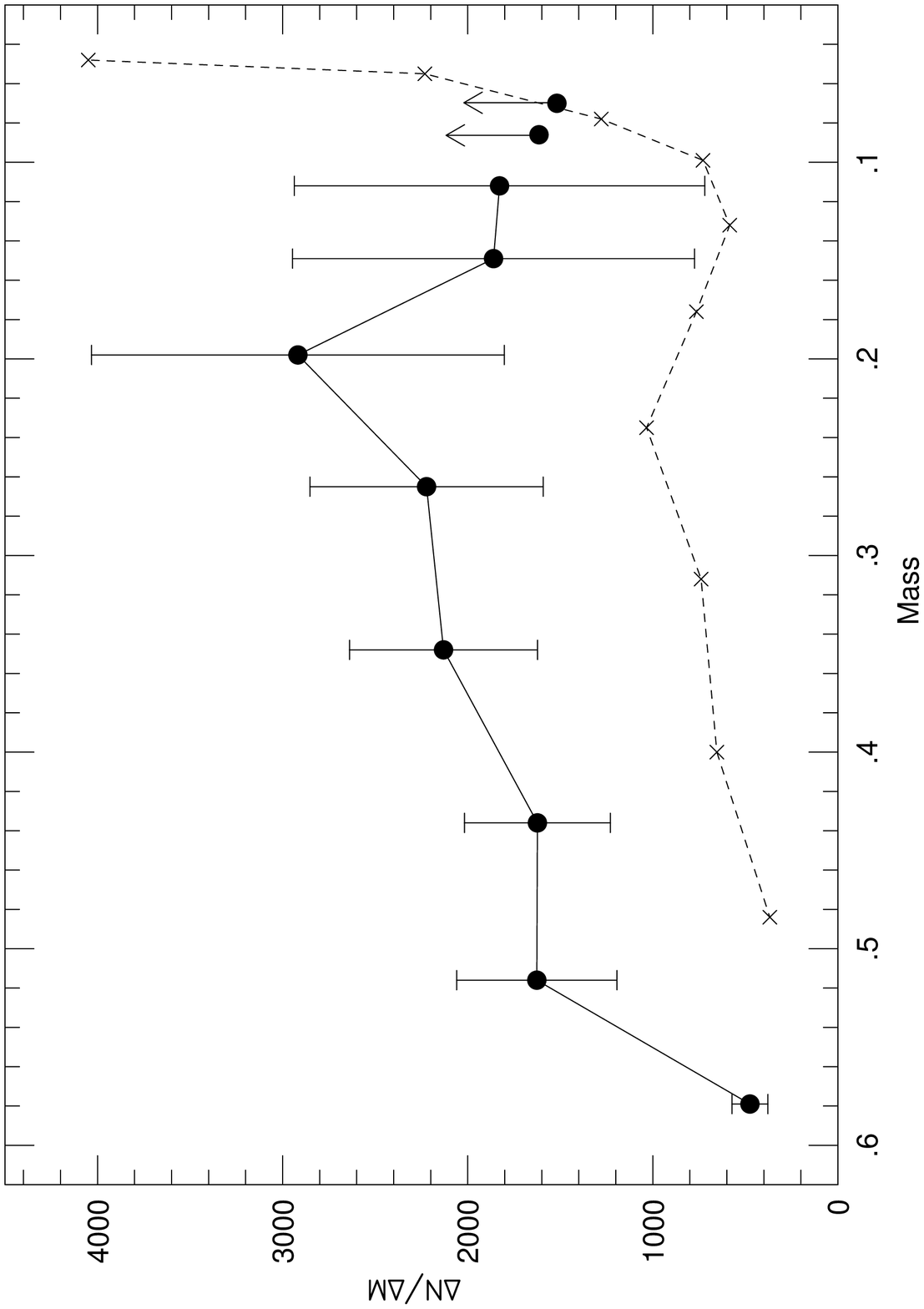}}
\vspace{0cm}
  \caption{
 Comparison between the luminosity functions ({\bf a})
 and the mass function ({\bf b}) of M35 and the Pleiades.}
\end{figure}

\section{The Luminosity Function}

We have subtracted the computed number of field stars for each
box from the measured number, for the boxes corresponding to the 
location of M35. The final value  is the number of  cluster members  and 
we have used it to compute the luminosity and initial mass functions.
 Note that we 
located the ZAMS in the diagram using the published values
in the literature and appropriate conversions to the colors we used.
We did not use this information when counting the number 
of stars in M35 and, therefore, 
our method is totally independent of distance modulus
and reddening. However, we have obtained that the maximum density of stars,
given a (R--I)$_{\rm c}$ value, appears $\sim$0.1$^{\rm mag}$ below 
the ZAMS, as a visual inspection indicates in Figure 1a. However, this could be
 due to the assumed distance and reddening values.

Figure 2a shows the Luminosity Function (LF) of M35, once the contamination 
by field stars is removed (solid line).  We have included
the errors as vertical bars.
 The distribution   peaks around I$_{\rm c}$=19. After this
point, the number of stars decreases with magnitude and a strong
contamination by field stars appears. Our sample is complete
until I$_{\rm c}$=21, (R--I)$_{\rm c}$=2.0.
 Therefore, the last two  points are lower limits.
 The M35 LF is quite similar to the one characteristic of the
Pleiades, shown as a dashed line in Figure 2a (Zapatero-Osorio,
1997). This distribution has been rescaled to match the values
of M35, since both studies cover different angular areas (1 sq.
deg. in the case of Pleiades against 0.25 sq. deg. for M35), and the
latter  cluster is more compact and is more populated (Leonard \&
Merritt, 1989). In any case, both distributions peak at similar
value, and present a sudden decrease. Contrary to what happens in
M35, the LF of the Pleiades increases below M$_{\rm I}$=11.
This difference is probably due to the fact that the Pleiades
study is much deeper, complete in the range, whereas this is not
true in our case.

\section{The Mass Function}

We estimated the masses using the I$_{\rm c}$ magnitudes. First,
we corrected for the reddening and the distance modulus,
obtaining M$_{\rm I}$ and (R--I)$_{\rm c,o}$ for each box.
Then, we convert the (R--I)$_{\rm c,o}$ colors into 
(V--I)$_{\rm c,o}$, using a polynomial expression between both
colors which was obtained by fitting the single nearby dM stars listed in
Leggett (1992). The (V--I)$_{\rm c,o}$  allowed us to obtain
bolometric corrections, using the conversion by Monet at al.
(1992) and, with them and M$_{\rm I}$, the bolometric
magnitudes. Finally, masses were derived from M(bol) and the
Baraffe (1997) isochrone, computed for an age of 120 Myr.
Figure 2b displays the number of stars per mass interval against
the mass. Data from M35 is shown as solid circles, whereas those
from the Pleiades  (Zapatero-Osorio 1997) appear,
 scaled for comparison purposes, as crosses.
 Errors bars are included for our data. 
The most important conclusion derived from this comparison is that
both IMF are identical. If the results in the BD domain 
are extrapolated to M35, this last cluster should contain
a significant population of BD. A more exhaustive analysis,
considering the mass range 6-0.1 M$_\odot$, is presented in 
Barrado y Navascu\'es et al. (1999).

\end{document}